\documentclass{aa}
\usepackage{graphicx}
\usepackage{natbib}

\begin{document}

\title{ Hot Very Small dust Grains in \object{NGC~1068} seen in  
jet induced structures thanks to  VLT/NACO adaptive optics
\thanks{ Based on observations collected at the ESO/Paranal YEPUN telescope,
Proposal 70.B-0307(A)}
}

\author{ 
	      D.\,Rouan \inst{1}
	\and  F.\,Lacombe \inst{1}
	\and  E.\,Gendron \inst{1}
    \and  D.\,Gratadour\inst{1,3}
    \and  Y.\,Cl\'enet  \inst{1} 
    \and  A.-M.\,Lagrange \inst{2}
    \and  D. Mouillet \inst{2}
    \and  C. Boisson \inst{5} 
    \and  G. Rousset \inst{3}
    \and  L. Mugnier \inst{3}
    \and  N. Thatte  \inst{4}
    \and  R. Genzel \inst{4}
    \and  P. Gigan  \inst{1}
    \and  R. Arsenault \inst{1,6}
    \and  P. Kern \inst{2}
}
    
\institute{ 
LESIA -- Observatoire de Paris-Meudon, UMR 8109 CNRS,  92195 Meudon,  France \\
 \email{daniel.rouan@obspm.fr}
\and  LAOG -- Observatoire de Grenoble,  UMR 5571 CNRS, 38041 Grenoble, France 
\and ONERA -- DOTA,   92322 Chatillon, France 
\and MPE,  Postfach 1312, 85741 Garching bei M\"unchen, Germany 
\and LUTH -- Observatoire de Paris-Meudon, UMR 8102 CNRS, 92195 Meudon, France 
\and ESO, Karl-Schwarzschild -Str. 2, 85748 Garching bei M\"unchen, Germany }
\offprints{D. Rouan}
\date{Received mar 15, 2003; accepted ....}

\abstract{ {\bf We present  K, L and M 
diffraction-limited images of \object{NGC~1068}, obtained with NAOS+CONICA at
VLT/YEPUN over a 3.5\arcsec $\times$ 3.5\arcsec
region around the central engine}.  Hot dust (T$_{col}$ = 550-650 K) is
found to be distributed in three main structurally different
 regions : 
(a) in the true nucleus, seen as a quasi-spherical,
however slightly NS elongated,  core of extremely hot dust,
{\it resolved} in K and L with respective diameters of 
$\approx$ 5 pc and 8.5 pc ;
(b) along the North-South direction, according to a spiral arm like
structure and a southern tongue ; 
(c) as a set of parallel elongated
nodules  ({\it wave-like}) on each side, albeit mainly at north, of
the jet, at a distance of  50 to 70 pc from the central engine.
The IR images reveal several  structures  also clearly observed  on
either radio maps, mid-IR or HST UV-visible maps, so
that a very precise registration of the respective emissions can
be done for the first time from UV to 6 cm. These results do
support the current interpretion that source (a) corresponds to emission
from dust near sublimation temperature delimiting the walls of the
cavity in the central obscuring torus. Structure (b) is thought to
be a mixture of hot dust and active star forming regions along a
micro spiral structure that could trace the tidal
mechanism  bringing matter to the central engine. Structure c)
which was not known, exhibits too high a temperature for
``classical'' grains ; it is most probably the signature of
transiently heated very small dust grains (VSG) : {\it nano-diamonds}, which
are resistant and can form in strong UV field or in shocks, are
very attractive candidates. The ``waves'' can be condensations
triggered by jet induced shocks, as predicted by recent models.
First estimates, based on a simple VSG model and on a detailed
radiative transfer model, do agree with  those interpretations,
both qualitatively and quantitatively.

 }
\titlerunning{Jet induced structures in \object{NGC~1068}}

\authorrunning{{\bf D. Rouan et al.}}

\maketitle

\section{Introduction}

 At a distance of 14.4 Mpc (70 pc per 1 \arcsec), \object{NGC~1068}
is unique for studying at the scale of a few pc the complex
immediate environment of an Active Galactic Nucleus (AGN). Indeed,
thanks to muti-wavelength studies, it has been possible in the
last decade to identify several distinct features : a structured
radio jet issued from a compact source, identified as the central
engine (\citealt{Gallimore96}), also seen in IR
(\citealt{Thatte97}) ; a structured molecular/dusty environment
around the central engine of \object{NGC~1068} detected at
infrared, millimetric and radio wavelengths
(\citealt{Gallimore96,Rouan98,Marco00,Schinnerer00,Gratadour03}) ;
a conical Narrow Line Region (NLR) seen at UV-visible wavelengths and
structured in high velocity ionized clouds (\citealt{Capetti95}).
Studying connections between those structures requires both a good
resolution and an excellent registration of the  maps at the
different wavelengths. Between radio and [visible + near-IR]
domains, thermal infrared, can play a unique role. Recently,
{\bf subarcsecond  imaging} in the L and M  bands has
been reported by \citet{Marco00} and \citet{Marco03} at a
resolution of $\approx$ 0.2 -- 0.5\arcsec. Here we report new
{\bf results at K, L and M where the spatial resolution and the sensitivity are pushed
further thanks to NACO (NAOS + CONICA) the new adaptive optics (AO) system of
the VLT, which offers, in addition to an excellent
correction of the atmospheric turbulence, the unique capability}
of thermal infrared imaging at a scale of 0.1\arcsec (\citealt{Rousset02,Lenzen02}).

\section{Observation and data reduction}

The observations were performed using NACO at the Nasmyth focus of
YEPUN,  during the nights 18-26 of
November  2003. The seeing was  good (typically O.6 \arcsec)
and NAOS was servoed on the bright nucleus itself, providing a
diffraction-limited correction, as proven for instance by the
three rings seen on the PSF in the M band. The  AutoJitter mode
was used, i.e. that at each exposure, the telescope  moves
according to a random pattern in a 6 \arcsec $\times$6 \arcsec
box. The pixel scale on CONICA is respectively 0.027 \arcsec 
 in the M and L bands  and 0.013 \arcsec in the K band. On
source total integration time was of 640, 800 and 448 sec at K, L
and M. A PSF reference  star, chosen to  give equivalent servoing
{\bf conditions of the AO system}, was observed just before
and after \object{NGC~1068}. Calibration files (flat field at dusk, dark exposures)
were acquired as ESO VLT standard data. The FWHM of the PSF,
estimated from reference stars was respectively 0.061\arcsec,
0.105\arcsec and 0.127\arcsec  at K, L and M.
Applied reduction procedures  are fully described in
Gratadour et al. (2003, in prep.). {\bf All images presented here are 
undeconvolved. }The dynamic on the final images
ranges from 770 at L to 3100 at M.

\begin{figure}
\centering
\includegraphics[width=8cm]{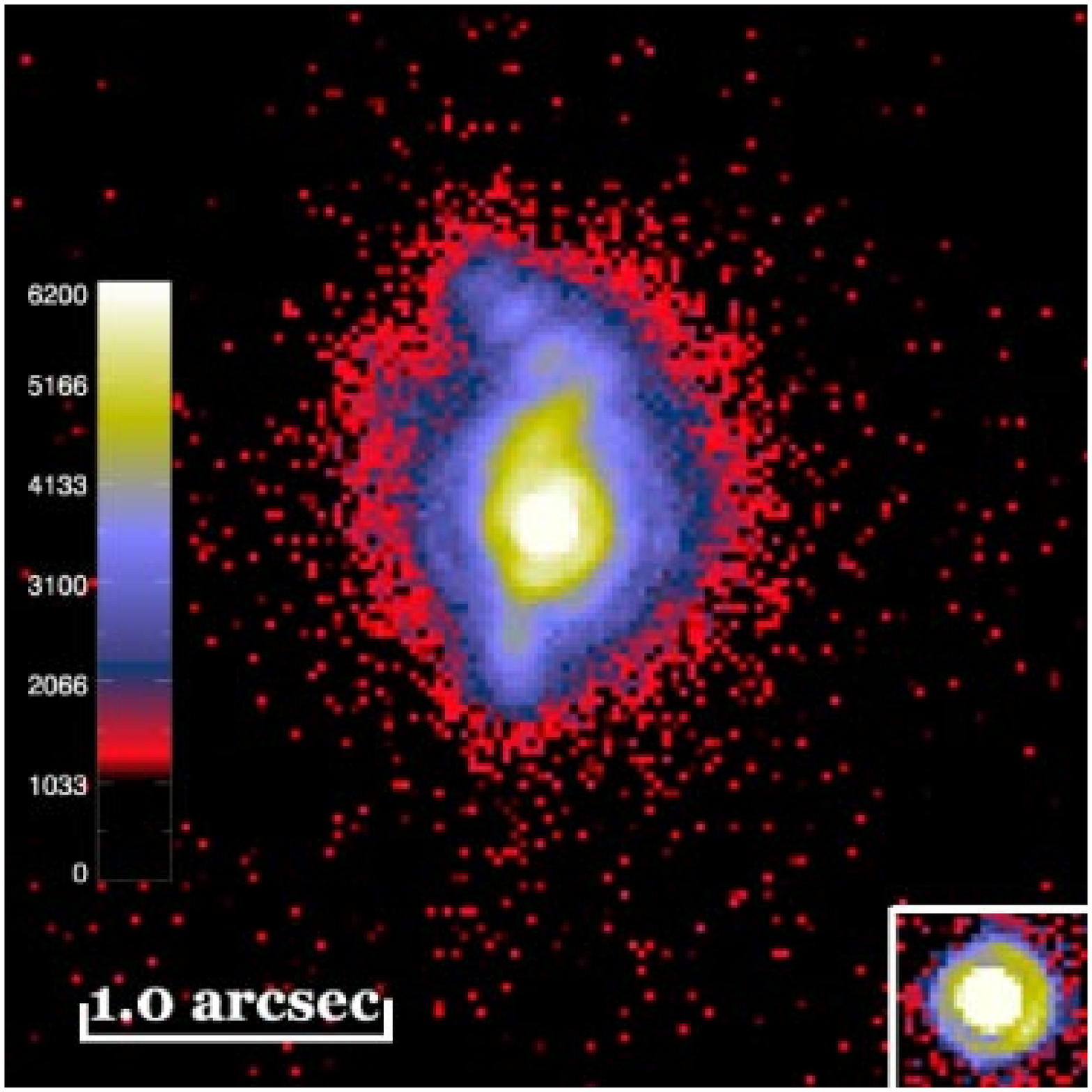}
\includegraphics[width=8cm]{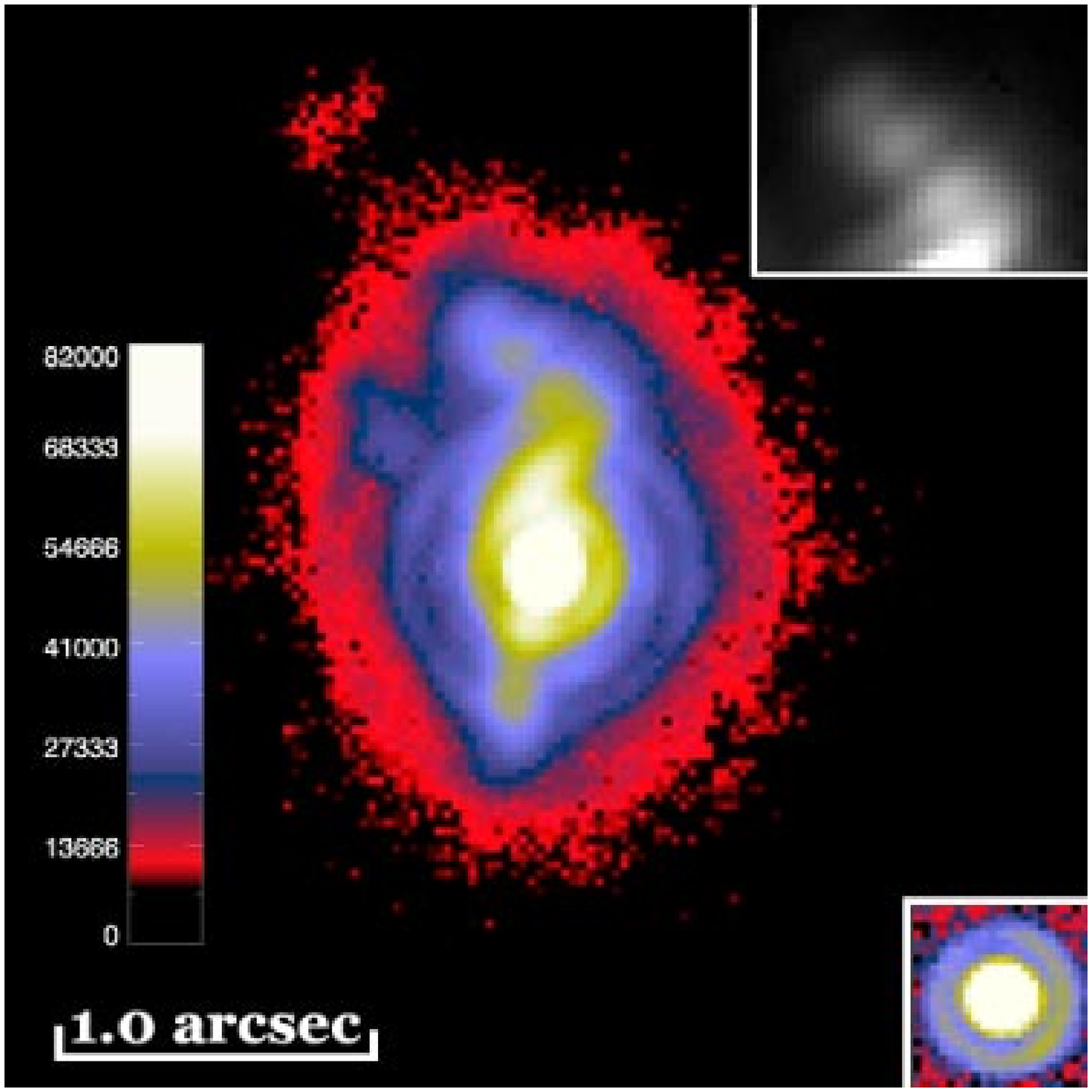}
\includegraphics[width=8cm]{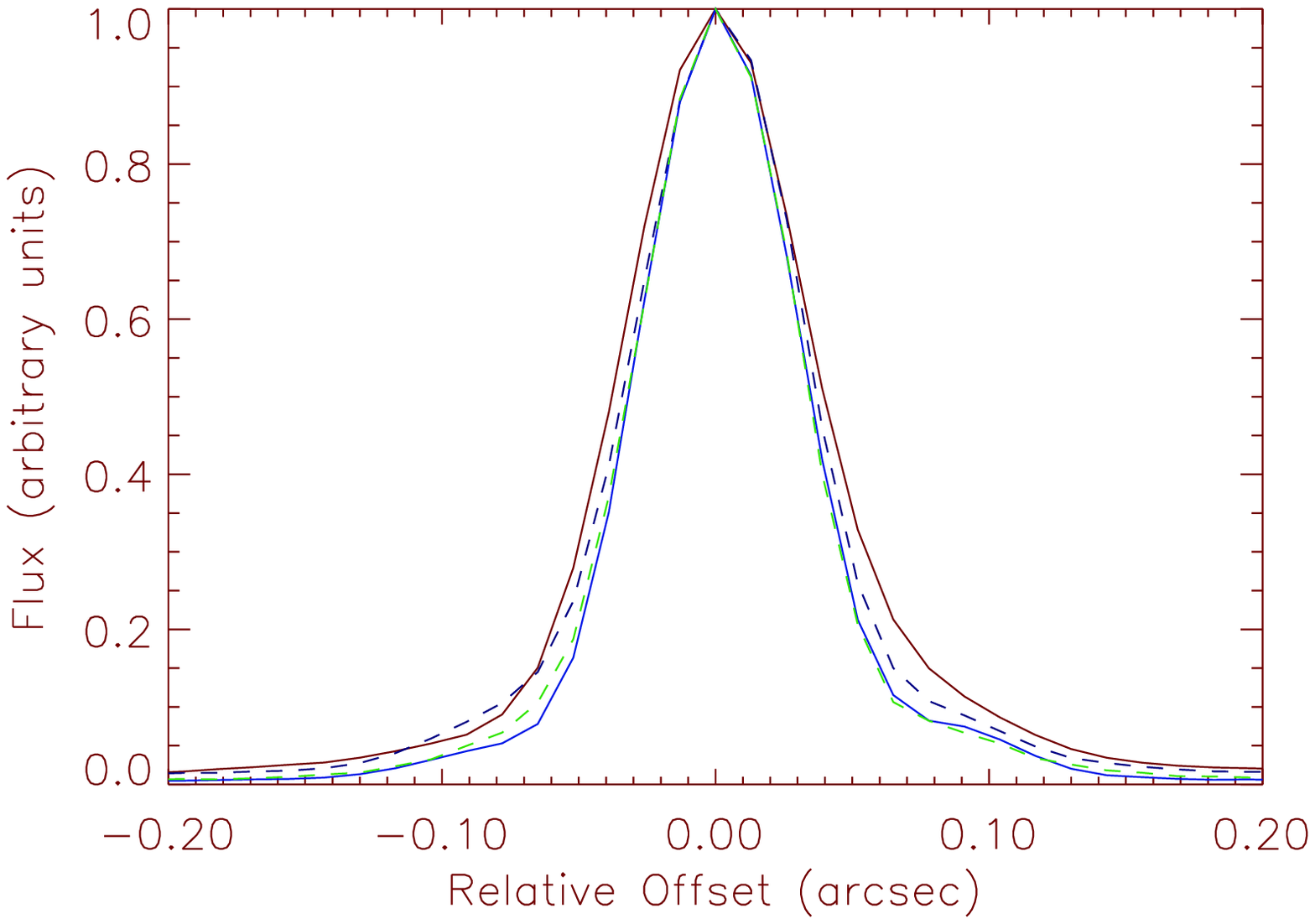}
\caption{ Top and middle : L and  M  band false-color images  around
the core of \object{NGC~1068}. The field is limited to 
3.46\arcsec $ \times$ 3.46\arcsec. The flux scale, indicated
on the color bar, is displayed on a power-law scale (I$^{0.8}$)
because of the high dynamic range
of the images.
The PSF is shown as an inset. Bottom
: cut of the core at K, along NS (solid brown line) and EW (dash yellow
line) axis ;  cut of the PSF along NS (solid blue line) and EW (dash green
line) axis. } \label
{fig1}
\end{figure}

\begin{figure}
\centering
\includegraphics[width=8cm]{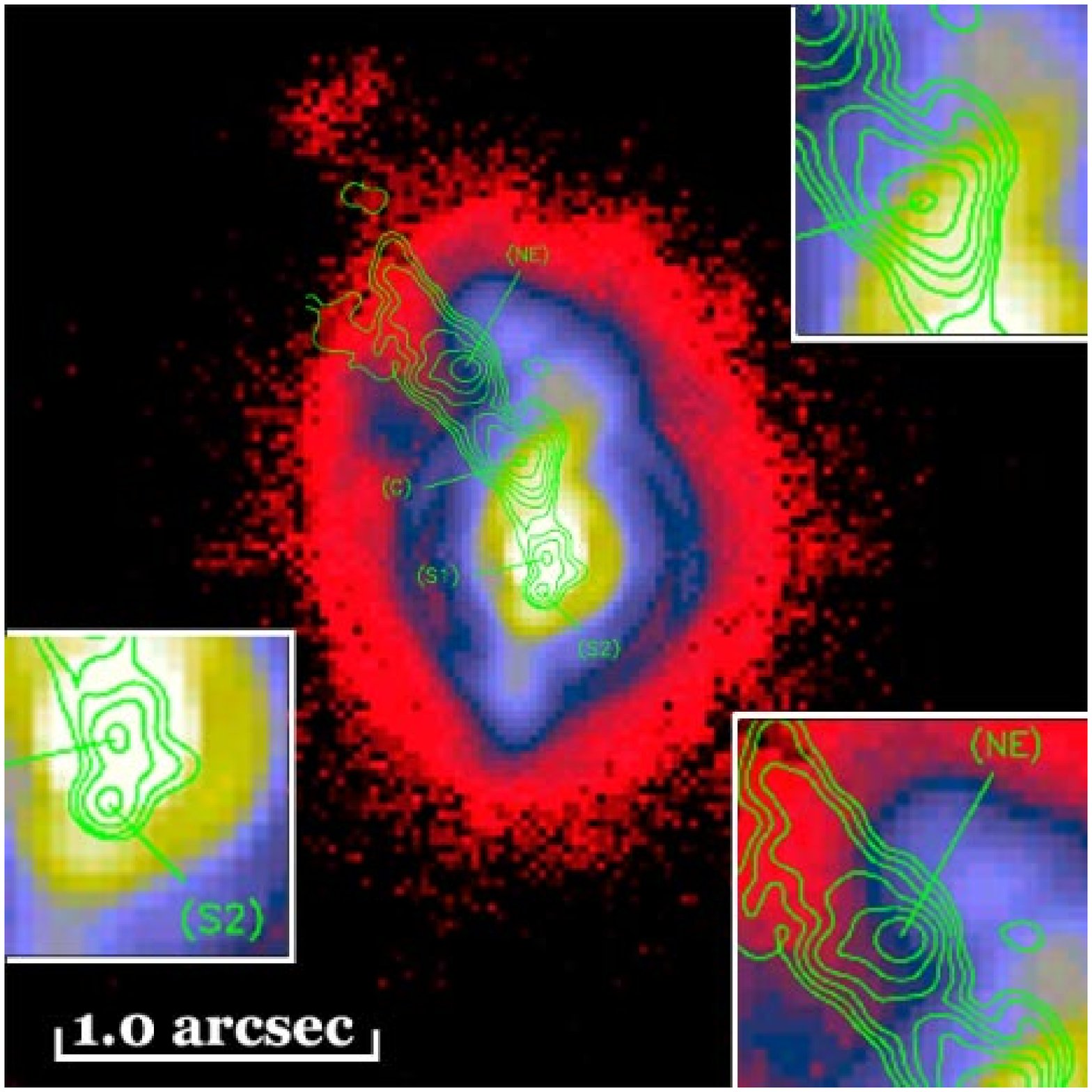}
\includegraphics[width=8cm]{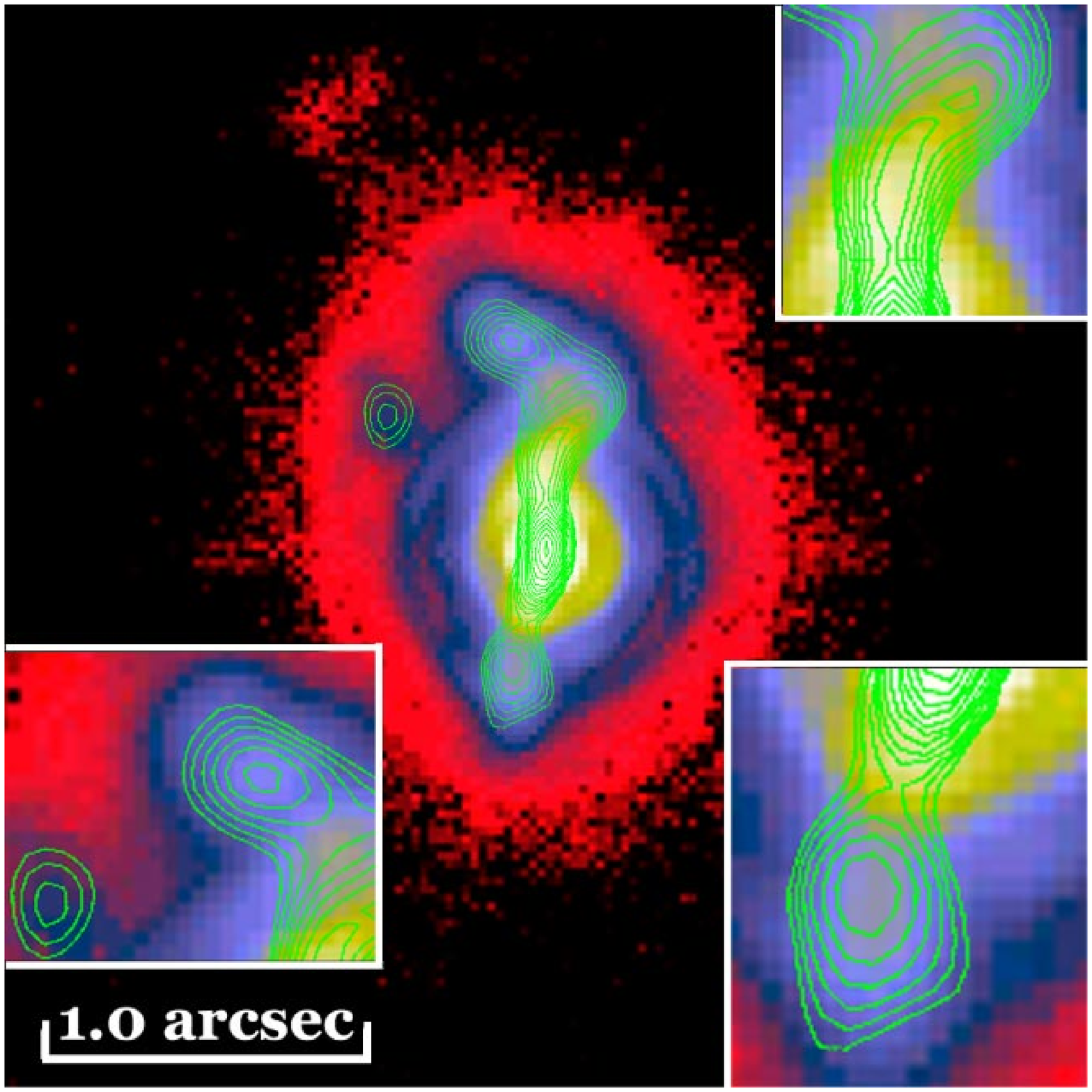}
\includegraphics[width=8cm]{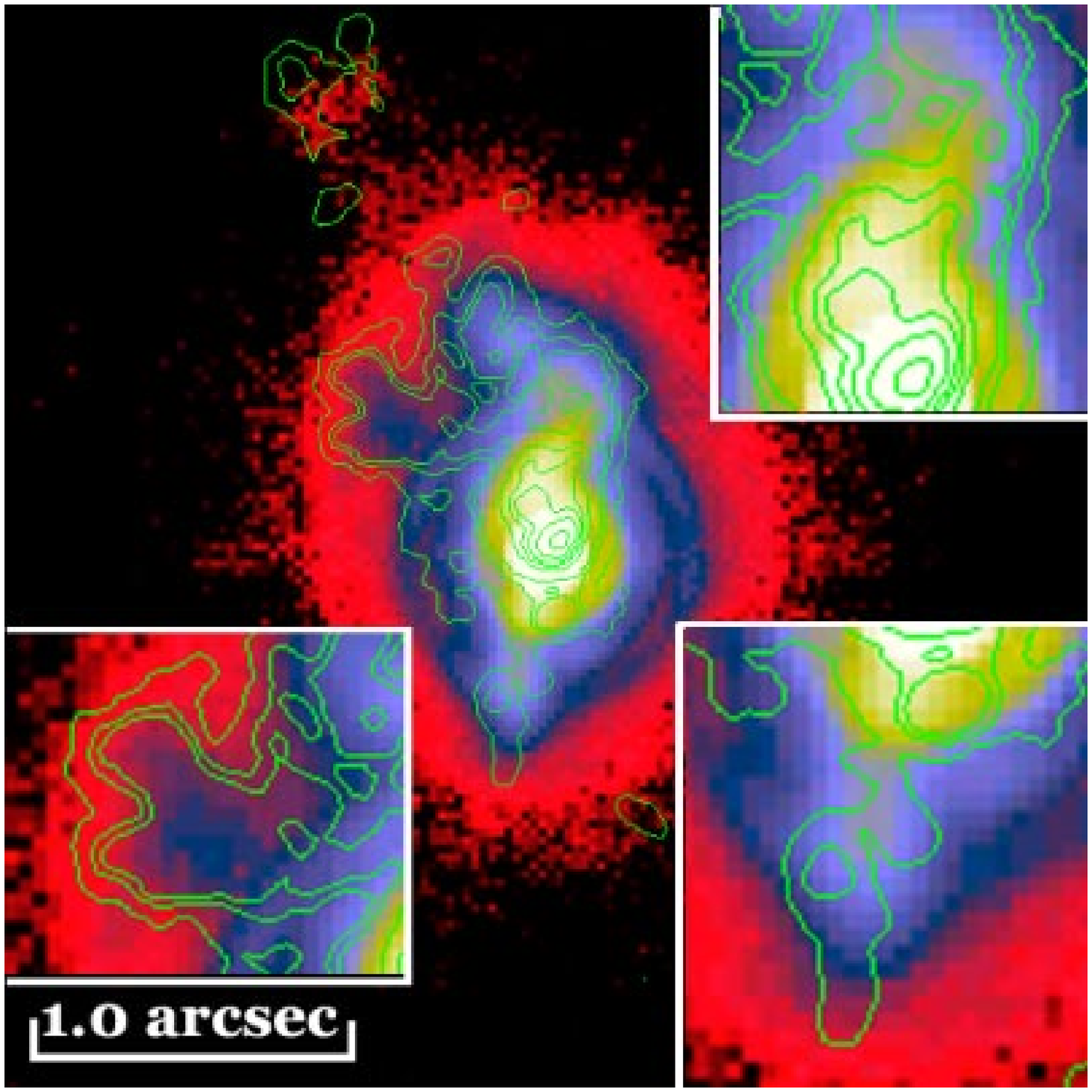}
\caption {  On our logscale M band image are superimposed
contours, from top to bottom, of : radio 5 GHz
(\citealt{Gallimore96}) ; 12.5 $\mu m$  (\citealt{Bock00}) ;
\ion{O}{iii} line emission (\citealt{Capetti95}). Various insets
of magnified regions are also shown. } \label{fig2}
\end{figure}

\section{Results and Discussion}

\subsection{Registration of radio IR and UV-visible images}

We show on Fig. 1,  images  at L and M, in a field  3.46\arcsec
$\times$ 3.46\arcsec.  The images are diplayed with a power-law
scale, so that {\bf details of the morphology} at all flux levels can be
seen. Since several noticeable features are found also on radio,
mid-IR and UV-visible images, a very precise registration
of all maps can be done. The most conspicuous structure on all
images is the bright central extremely compact source, already
known at those wavelengths as well as at 10 $\mu m$. {\bf The bright IR source
is most likely the counterpart of the  5 GHz radio source S1 that is
unambiguously} identified with the central
engine because of its spectrum and of the distribution of the
maser sources around it (\citealt{Gallimore01}). The mid-IR images
(\citealt{Bock00}) also show a bright central core which obviously
marks the location of the same
 source. {\bf The contours of the radio and the mid-IR
emissions can thus be superimposed precisely on our IR image, as 
shown on Fig. 2a-b}. A much
fainter feature, only present  on the M image, is an
elongated patch 2.1\arcsec NNE from the core, while another one,
seen both at L and M, is a tongue at 0.69\arcsec NE, with a
characteristic shape of a crab claw. Those two features are
clearly seen on the images in the UV-visible domain
(\citealt{Capetti95}), {\bf so that it is possible to  locate
unambiguously -- within $\approx$ 0.05 \arcsec -- the IR
core on the visible image.} Our
cross-identification (Fig. 2c) shows that the true nucleus is in
fact at the apex of the UV cone, i.e. coincident with the
center of polarization vectors in the near-IR and mid-IR
(\citealt{Lumsden99}) : this is the location adopted by
\citet{Alloin01} {\bf and more recently by \citet{Galliano03}
who made a very precise astrometric registration of the UV, radio and K maps. 
Two important correlations are also
seen on Fig. 2 }: a) the radio jet is bordered in L and M by
two extended tongues parallel to the jet (the northern being more
conspicuous), with enhancements that are tightly correlated ; b)
the 12 $\mu m$ map does correlate extremely well with the
brightest features of the M and L maps : the NS extension, the NE
tongue at the end of the extension and the NE ``crab claw''.

\subsection{The size of the central cavity}

The most favoured  interpretation (\citealt{Thatte97}) is that the
core corresponds to the emission  from the (UV heated) very hot
dust -- close to sublimation temperature (1000-1500 K) --
delimiting the wall of the central cavity around the accretion
disk. This interpretation was recently fully confirmed by CFHT
diffraction-limited spectroscopy at a spectral resolution of 220
(\citealt{Gratadour03} : GRCLF03), that allowed to derive a color
temperature of 950 K and a deredenned one of 1200 K, i.e. close to
sublimation temperature of silicates. The new fact brougth by our
observations is that the emission is clearly resolved at K and
at L, as revealed by cuts across the core displayed on
Fig. 1-c. The emission is especially well resolved in the
north-south direction, where the measured FWHM is .067\arcsec at K 
and .122\arcsec at L, i.e. 4.7 pc and 8.5 pc
respectively. The FWHM at L is larger than at K because it comes
from somewhat cooler, more extended dust. When comparing those
results to our model (GRCLF03) -- which fairly describes the continuum spectrum
in the K band at a scale of 0.15\arcsec--, the core size at K and L, 
the NS extension, as well as the {\bf observed} flux ratio are
reproduced to within 10 $\%$. {\it The fact that the cavity is resolved is
thus a very strong support of the unified AGN scheme}.

\subsection{Micro-spiral like structure}

Two structures close  to the AGN are well defined, especially at
M. The same structure as depicted by \citet{Rouan98} of a spiral
arm beginning sligthly NE from the core and bending clockwise, up
to a point at $\approx$ 0.45\arcsec (30pc) NNW from the center,
is observed at all wavelengths. Similarly, a tongue to the south
(at a P.A. of 86$\degr$) is clearly observed, with the
difference that it does not show a spiral-like structure ; it
may be obscured by dust from the putative tilted dusty torus (GRCLF03).
In both cases, the M/L flux ratio of 1.8 implies a rather high
color temperature, larger than 600K. This ratio is  correctly
reproduced by thermal emission of  grains at 475 K with  Q$_{abs}
\propto \lambda^{-1.5}$. At this distance (20 pc) from the AGN,
the estimated temperature of classical grains with this emissivity
is indeed around 475 K, assuming a direct heating by the hard
radiation from the accretion disk.  In any cases, even an
extremely dense stellar cluster cannot heat dust so efficiently.
The observed spiral structure is thus more likely dust heated by
the central source. It could well traces the innermost
stage of nested spirals systems : this type of structure was
proposed to produce the required braking torque that brings matter
to the center to ultimately feed  the ``monster''
(\citealt{Shlosman89}).

\section{Hot Very Small Grains in wave-like structures around the jet}

This is probably the most striking result we obtained : to the
north of the radio jet (see Fig. 2a), four parallel elongated
nodules of $\approx$ 0.2\arcsec long each, forming a kind of
wave pattern, are observed mainly at M and L, but also at a much
fainter level at K. This is especially clear on the inset of Fig.
1-a. The first of those structures is in fact the end of the
spiral arm described in the previous section. The three others are
fairly parallel to it and distributed along a direction which has
the same P.A. as the jet, as if they were delineating the cocoon
of the jet. Indeed, the mid-IR images from Bock et al. show an
elongated tongue that superimposes very well on the set of "waves"
we observe. However, no clues of a sub-structuration were seen on
the 12.5 $\mu m$ deconvolved image. The K image, despite the low
S/N, indicates that those structures are intrinsically very narrow
and probably unresolved in the direction of the jet. It is
unlikely that they correspond to classical clouds of gas and/or
dust that would be aligned by chance, unless we just see  their
illuminated front. If the {\bf pseudo-periodicity} corresponds actually to some
physical phenomenon, then the wavelength would be typically of 10
pc. {\bf We rather favour the interpretation that this scale  is typical of
instabilities} in the ISM, which developped because of the compression
by the jet cocoon. Hydrodynamical models of this interaction
indeed predict a nested cylindrical structure with dense clumps on
each side of the jet corridor (\citealt{Steffen97}, Fig. 2b). The
photometry in L and M reveals unexpectedly high temperatures :
for instance, a color temperature of 550 K is derived for the
third, well delineated, nodule. At a projected distance of 70 pc,
the energy density that is expected at the level of the nodule,
assuming that there is no screening of the UV, is $\approx$ 10$^5$
eV. Even under such a large irradiation, the temperature of a {\it
classical} grain would only be 330 K. Redenned free-free emission
could be a possible mechanism in a region where the gas is
essentially ionized, since the flux density is about twice at M
than at L. However, the corresponding radio  or mid-IR emission
would be respectively much higher and much lower than observed.
Moreover, \citet{Bock00} have shown that the radio jet cannot be
the only heating source, unless the conversion  of  the
shock energy into IR emission is extremely efficient. 
{\bf To reach high temperatures, heating by 
soft X-rays is a more  plausible mechanism 
that must be explored. Another appealing hypothesis
that we favour, is that transiently heated very small grains
(VSG) are responsible for this emission}. Indeed, this component of
the interstellar dust has been invoked years ago
(\citealt{Sellgren84}) and the prediction that it can contribute
very significantly in the near to mid-IR range when the UV field
is very strong, is not recent (\citealt{Desert90}). Recently, this
mechanism was  proposed to explain  the very red colors observed
in ULIRGs (\citealt{Davies02}). Nanodiamonds are in fact very good
candidates for those VSG in the case of an AGN : (a) they can 
form very efficiently in a strong UV field or in shocks
(\citealt{Jones00}) ; (b) they are not easily destroyed and (c) they
can reproduce very well the observed color temperature. 
Using Jones and d'Hendecourt (2000) heat
capacity, we built a model of transient heating,  and computed that, 
for instance, a  0.5 nm diamond
absorbing a 6 eV photon and emitting in the IR while cooling,
reproduces the observed L/M ratio of nodule \#3 ; a mixture of
nanodiamonds from 0.5 to 2 nm receiving photons from 1 to 7.5 eV
reproduces the ratio of L, M {\it and} N  (Bock et al., 2000) fluxes to within 8 $\%$.
More refined fits will soon be presented (Gratadour et al., in
prep.). We predict that several characteristic lines of nanodiamonds
should be detectable in this range.

\acknowledgements
We thank the ESO  team on Paranal, particularly
O. Marco, for the support during observations
and the NAOS and CONICA consortia for their superb instruments. 
Help from C. Catalano was also deeply appreciated.

\end{document}